\documentclass[prl,amsmath,amssymb,aps,superscriptaddress,twocolumn,nofootinbib
]{revtex4-2}
\DeclareUnicodeCharacter{2299}{\odot}

\usepackage{epsfig}
\usepackage{url}
\usepackage{hyperref}

\usepackage[normalem]{ulem}

\usepackage{latexsym}
\usepackage{epsfig}
\usepackage{amsmath}
\usepackage{amssymb}
\usepackage{wasysym}
\usepackage{graphicx}
\usepackage{dcolumn}
\usepackage{verbatim}
\usepackage{enumerate,mdwlist}

\usepackage[titletoc]{appendix}

\usepackage{amsfonts}
\usepackage{fancyvrb} 
\usepackage{tikz} 
\usetikzlibrary{calc}
\usepackage[export]{adjustbox}

\usepackage[normalem]{ulem}

\usepackage[inline, final]{showlabels}

\usepackage{listings}
\usepackage{xcolor}

\definecolor{comment_red}{rgb}{0.5, 0, 0}
\makeatletter
\g@addto@macro\bfseries{\boldmath}  
\makeatother

\newcommand{\ba}{\begin{eqnarray}}
\newcommand{\ea}{\end{eqnarray}}

\definecolor{grey}{rgb}{0.4,0.4,0.4}
\definecolor{dullmagenta}{rgb}{0.4,0,0.4}
\definecolor{darkblue}{rgb}{0,0,0.4}
\definecolor{midblue}{rgb}{0,0,0.5}
\definecolor{midred}{rgb}{0.5,0,0}
\definecolor{orange}{rgb}{1,0.5,0}
\definecolor{lightbrown}{rgb}{0.75,0.5,0.25}
\definecolor{tan}{cmyk}{0.14,0.42,0.56,0}
\definecolor{djunglegreen}{cmyk}{0.99,0,0.52,0}
\definecolor{lightgreen}{rgb}{0,1,0}
\definecolor{olivegreen}{cmyk}{0.64,0,0.95,0.40}
\definecolor{midgreen}{rgb}{0.0,0.675,0.0}
\definecolor{darkgreen}{rgb}{0,0.5,0}
\definecolor{ceruleanblue}{rgb}{0.0, 0.2, 0.7}
\definecolor{burgundy}{rgb}{0.5, 0.0, 0.13}
\definecolor{hvred}{RGB}{186,12,47}

\newcommand{\mz}[1]{\textcolor{black}{#1}}
\newcommand{\shan}[1]{\textcolor{black}{#1}}

\hypersetup{
    colorlinks=true,
    linkcolor=ceruleanblue,
    filecolor=ceruleanblue,
    urlcolor=midblue,
    citecolor=burgundy,
}

\usepackage{booktabs}

\usepackage[all]{xy} 
\usepackage{amsfonts}

\makeatletter
\def\l@subsubsection#1#2{}
\makeatother

\begin{document}


\title{Gravitational lensing of gravitational waves by galaxy clusters}


\author{Xikai Shan}
\email{xk\_shan@mail.bnu.edu.cn}
\affiliation{Department of Astronomy, Tsinghua University, Beijing 100084, China}
\affiliation{Max Planck Institute for Gravitational Physics (Albert Einstein Institute) \\
Am Mühlenberg 1, D-14476 Potsdam-Golm, Germany}

\author{Yi-Fan Wang}
\email{yifan.wang@aei.mpg.de}
\affiliation{Max Planck Institute for Gravitational Physics (Albert Einstein Institute) \\
Am Mühlenberg 1, D-14476 Potsdam-Golm, Germany}
\affiliation{Purple Mountain Observatory, Chinese Academy of Sciences, Nanjing 210034, China}

\author{Shude Mao}
\email{shude.mao@westlake.edu.cn}
\affiliation{Department of Astronomy, Westlake University, Hangzhou 310030, Zhejiang Province, China}

\author{Huan Yang}
\email{hyangdoa@tsinghua.edu.cn}
\affiliation{Department of Astronomy, Tsinghua University, Beijing 100084, China}

\author{Miguel Zumalac\'arregui}
\email{miguel.zumalacarregui@aei.mpg.de}
\affiliation{Max Planck Institute for Gravitational Physics (Albert Einstein Institute) \\
Am Mühlenberg 1, D-14476 Potsdam-Golm, Germany}
\affiliation{Instituto de Física Teórica UAM/CSIC, \\
C/ Nicolás Cabrera 13--15, Universidad Autónoma de Madrid, \\
Cantoblanco, 28049 Madrid, Spain}



\begin{abstract}
Strongly lensed gravitational waves (GWs) are commonly searched for using population priors, including time-delay and magnification-ratio distributions, to suppress false alarms. Current LIGO, Virgo, and KAGRA (LVK) searches mainly adopt priors calibrated with galaxy-scale lenses; however, although these priors effectively reduce accidental event pairs, they may penalize the longer time delays produced by galaxy clusters, which could contribute an appreciable population of lensed events (roughly 10\% to 20\% of the entire lensing sample). Moreover, cluster member galaxies may perturb the macro-lens potential and alter image magnifications. In this work, we investigate the time-delay and magnification-ratio distributions of binary black hole (BBH) GWs lensed by galaxy clusters, together with their microlensing systematics. In the full lensed sample, the median time delay is about $76$ days for cluster lenses, compared with $6.9$ days for galaxy lenses. After imposing an SNR threshold of $\rho>8$ at LVK O4 sensitivity, the corresponding medians are $15.4$ and $0.24$ days, respectively. Furthermore, the detected cluster sample retains a long-delay tail, with $9.1\%$ of image pairs delayed by more than $365$ days and $3.3\%$ by more than $1000$ days. Consequently, galaxy-based time-delay priors can artificially suppress the selection probability of cluster-lensed image pairs. By contrast, although cluster member galaxies broaden the magnification-ratio distribution of the full cluster-lensing sample, the detected cluster- and galaxy-lensing samples have very similar distributions at O4 sensitivity, suggesting that differences in magnification ratio are unlikely to substantially bias candidate selection. In addition, cluster lenses generally produce weak microlensing distortions: only about $0.8\%$ of single-image mismatches exceed the matched-filter waveform-accuracy benchmark ${\cal M}=0.03$, while $4.8\%$ exceed the parameter-estimation threshold $1.35/\rho^2$ at O4 sensitivity. Therefore, microlensing is a subdominant waveform systematic for most cluster-lensed events and can generally be neglected in searches for cluster-lensing signatures. This is mainly because images often form far from the brightest cluster galaxy, where the stellar density is low and dominated by intracluster light, although nearby satellite galaxies can locally enhance the stellar density. Taken together, these results demonstrate that the distinct time-delay distribution and complex lensing environments of galaxy clusters motivate a dedicated cluster-lensing pipeline for GW lensing searches.

\end{abstract}

\date{\today}

\maketitle

\tableofcontents

\section{Introduction}
Strong gravitational lensing of GWs provides a direct way to study a wide range of topics in cosmology and astrophysics, including time-delay cosmography to constrain the Hubble constant, tests of GW propagation beyond general relativity, studies of \mz{dark-matter properties}, constraints on the high-redshift merger population, and localization of the host galaxy of the GW source~\citep{Sereno2011,Liao2017,Hannuksela2020,Oguri2020,Xu2022,Ng:2024ooy,Narola:2023viz,Janquart:2023mvf,Wempe:2022zlk,Ezquiaga_2020,Goyal:2023uvm,Urrutia:2021qak,Tambalo:2022wlm,GilChoi:2023ahp,Cheung:2024ugg,Smith:2022vbp,Smith:2025axx,Shan:2023qvd,Seo:2025dto,Shan:2025jpt}.
However, identifying strongly lensed GW events with high efficiency and a low false alarm rate remains an active problem~\citep{Caliskan:2022wbh,Hannuksela:2025wgv,Barsode:2025agk,Smith:2022vbp}.
The most robust and widely used methods comprise an astrophysical-prior-weighted pipeline that progresses from rapid posterior-overlap, reconstructed-phase-consistency, or waveform-consistency tests to more precise joint parameter estimation
~\citep{Haris2018,Wright:2025nod,Janquart:2023osz,Janquart:2022wxc,Lo2021,Ezquiaga:2023xfe,More_2022,Goyal:2021hxv,o4alensing,Abbott2024Lensing,Abbott2021Lensing,Hannuksela2019}.

However, as the GW catalogue grows, the number of random event pairs increases rapidly~\citep{Caliskan:2022wbh}.
Some unrelated pairs can have overlapping posterior distributions by chance, especially when the events have broad parameter uncertainties.
Therefore, in this method, astrophysical priors, such as the time-delay and magnification-ratio priors, are key to its success because parameter consistency alone has a false-alarm rate that is too high to identify a secure lensed pair~\citep{Haris2018,Barsode:2025agk,Wierda_2021,More_2022,Janquart_2022}.
This raises another issue: to avoid incorrectly ruling out genuine strongly lensed pairs or incorrectly accepting false alarms, the astrophysical prior must be as realistic as possible.


Recently, the limitations of such priors have also received increasing attention. For example, \citet{Barsode:2026yqc} used a time-delay and magnification-ratio prior predicted by singular isothermal ellipsoid (SIE) models~\citep{1994A&A...284..285K} covering a cluster-scale mass range to screen strongly lensed pairs across O1, O2, O3, and O4a. Nevertheless, to date, the time-delay and magnification-ratio priors used in most studies have been calibrated primarily using galaxy-scale lensing models, and no study has used a prior predicted from a more realistic galaxy-, group-, and cluster-lensing population, such as those of~\citet{Robertson:2020mfh, Li:2026dai}.
This choice can affect searches for events with long time delays, such as event pairs spanning O1, O2, O3, and O4a, because the full observing baseline is much longer than the typical time delays expected from galaxy-scale lenses. Moreover, substructures within galaxy clusters may also affect the predicted magnification-ratio distribution~\citep{Vujeva:2025nwg,Vujeva2025}.
In this study, we use end-to-end simulations to investigate two macro-lensing observables, namely the time delay and magnification-ratio between image pairs, as well as the microlensing-induced waveform mismatch in cluster-lensed GWs.

The first question is the time-delay distribution.
We ask whether the image-pair delays of GWs lensed by clusters are much longer than those of GWs lensed by galaxies~\citep{Smith_2018}.
If so, searches that use only a galaxy-lensing time-delay prior will miss part of the population lensed by clusters.
In that case, cross-run searches and a cluster-lensing time-delay prior are needed when selecting strong-lensing candidates.

The second question concerns the magnification-ratio distribution. Cluster member galaxies and dark-matter substructures can perturb the macro-image magnifications and may broaden the magnification-ratio distribution relative to that predicted by smooth galaxy-scale lens models~\citep{Vujeva2025,Vujeva:2025nwg}. We therefore ask whether a magnification-ratio prior calibrated using galaxy-scale lenses could also bias the selection of cluster-lensed image pairs, particularly at O4 sensitivity.

The third question concerns waveform distortions induced by microlensing.
This effect may differ from that in galaxy-scale lenses because many cluster macro images form far from the BCG.
At these positions, the local compact-object convergence is mainly associated with the diffuse ICL, which has a relatively low stellar mass surface density.
For example, \citet{Venumadhav_2017} estimated that the intracluster stellar component can contribute a convergence of order $\kappa_\star\sim10^{-2}$ near the critical curves of a cluster lens.
In addition to the ICL component, substructures such as satellite galaxies within the cluster can also affect the image multiplicity, magnification~\citep{Vujeva2025,Vujeva:2025nwg}, and local microlensing density at macro-image positions.
The latter two quantities can further affect the microlensing signatures of cluster-lensed GWs.
This multi-component nature makes cluster microlensing different from galaxy-scale microlensing and makes it an interesting problem in its own right.


Before undertaking this study, the first important question to address is the GW lensing rate from clusters.
\citet{Chen2024} showed that, for group- and cluster-scale halos in the mass range $10^{13}M_\odot < M < 10^{15}M_\odot$, the lensing optical depth is only somewhat smaller than that of galaxy lensing.
In the fiducial estimate of \citet{Chen2024}, which uses CosmoDC2 mock clusters~\citep{2019ApJS..245...26K}, the detection rates of group/cluster-lensed stellar BBH mergers by third-generation GW detectors are $13^{+28}_{-2}\ {\rm yr^{-1}}$ and $19^{+5}_{-13}\ {\rm yr^{-1}}$ for the isolated binary and dynamical formation channels, respectively. These values are comparable to the corresponding galaxy-lensing rates of $14^{+30}_{-2}\ {\rm yr^{-1}}$ and $23^{+7}_{-16} \ {\rm yr^{-1}}$ \citep{chen2023,Chen2024}. In addition, \citet{Chen2024} also found that the contribution from halos with $10^{13}M_\odot < M < 10^{14}M_\odot$ (the group range) is about $1.5$--$4$ times larger than that from halos with $10^{14}M_\odot < M < 10^{15}M_\odot$ (the cluster range). Therefore, if only cluster-scale halos with $10^{14}M_\odot < M < 10^{15}M_\odot$ are considered, the lensing rate would be reduced to roughly $2.6$--$5.2 \ {\rm yr^{-1}}$ and $3.8$--$7.6 \ {\rm yr^{-1}}$ for the two channels. In this case, the contribution from $10^{14}$--$10^{15}M_\odot$ cluster lenses would account for $\sim 10$--$20\%$ of the total galaxy plus group/cluster lensing detections.
This is encouraging for studies and searches of cluster-lensing events.

In addition, strongly lensed supernovae provide electromagnetic evidence that cluster-scale lenses are an important transient lensing channel.
To date, the published sample of multiply imaged lensed supernovae contains at least nine well-studied systems.
Several of them are produced by galaxy clusters or cluster fields, including SN Refsdal, SN Requiem, SN H0pe, SN Encore, the red-supergiant supernova at $z\simeq3$ behind Abell 370, and SN Eos \citep{Kelly2015,Rodney2021,Pascale2025,Pierel2025,Chen2023SN,Coulter2026}.
The isolated galaxy-scale systems include iPTF16geu, SN Zwicky, and SN 2025wny \citep{Goobar2017,Goobar2023,Taubenberger2025}.
\shan{Although the observed supernova sample is still small and subject to selection bias because galaxy clusters are actively monitored, it is already sufficient to demonstrate that cluster lenses can produce detectable multiply imaged transients, thereby motivating dedicated searches for GWs lensed by clusters.}


The rest of this paper is organized as follows.
In Section~\ref{sec:methods}, we describe the GW source population, the cluster lens population, the macro lens model, and the microlensing calculation.
In Section~\ref{sec:results}, we present the time-delay and magnification-ratio distributions of cluster-lensed image pairs, as well as the microlensing-induced waveform mismatch of cluster-lensed GWs.
In Section~\ref{sec:conclusion}, we discuss the implications for future cluster lensing searches in GW catalogues.

\section{Methods}\label{sec:methods}
In this section, we describe the method used to simulate BBH GW events lensed by clusters.
The mock-data pipeline has five main steps.
We first generate the intrinsic BBH population.
We then assign cluster lenses through the strong-lensing optical depth, construct cluster lens realizations including the smooth halo, BCG, ICL, satellite galaxies, and subhalos, solve the macro lens equation, estimate the stellar convergence at each macro image, and compute the macro-lensed and microlensed waveforms.
The Monte Carlo strategy follows the strongly lensed GW simulations in \citet{Shan2025}.
The smooth cluster halo follows the elliptical Navarro--Frenk--White (NFW) framework used in \citet{Chen2024}, with the BCG, ICL, satellite-galaxy, and subhalo components added as described below.
More complex lens configurations containing multiple cluster-scale mass components, such as those illustrated by \citet{2024ApJ...974...23V}, are not considered in this work.

\subsection{BBH population}

The source redshift is sampled from a merger-rate model based on the star formation rate density of Madau and Dickinson \citep{MadauDickinson2014},
\begin{equation}
\rho_{\rm SFR}(z)=
0.015\,
\frac{(1+z)^{2.7}}
{1+\left[(1+z)/2.9\right]^{5.6}}
\,{\rm M_\odot\,yr^{-1}\,Mpc^{-3}} .
\end{equation}
Following the treatment used in lensed GW population simulations \citep{Haris2018,Xu2022,Shan2025}, we assume that the merger rate traces the past star formation history with a time delay.
The delay-time distribution is taken to be $p(t_d)\propto t_d^{-1}$, with a minimum delay time $t_{d,\min}=50\,{\rm Myr}$.
For a binary that merges at redshift $z_s$, the delay time is
\begin{equation}
t_d=t_{\rm lb}(z_f)-t_{\rm lb}(z_s),
\end{equation}
where $z_f$ is the formation redshift and $t_{\rm lb}$ is the lookback time.
The merger rate is then defined as
\begin{equation}
\begin{split}
\mathcal{R}(z_s)
\propto
&\int_{z_{f,\min}}^{z_{\rm max}}
\rho_{\rm SFR}(z_f)
p(t_d)
\frac{dt_{\rm lb}}{dz_f}
dz_f \\
\propto&\int_{z_{f,\min}}^{z_{\rm max}}
\frac{\rho_{\rm SFR}(z_f)}{(1+z_f)}
p(t_d)
\frac{1}{H(z_f)}
dz_f.
\end{split}
\end{equation}
Here $z_{f,\min}$ is set by $t_d=t_{d,\min} (50\,{\rm Myr})$, and $z_{\rm max}=1100$ in the numerical calculation.

The source redshift distribution is sampled from
\begin{equation}
\frac{dN}{dz_s}
=
\frac{\mathcal{R}(z_s)}{1+z_s}\,
4\pi
\frac{c\,D_c^2(z_s)}{H(z_s)} ,
\end{equation}
where $D_c(z_s)$ is the comoving distance and $H(z_s)$ is the Hubble parameter.

For the BBH parameters, we follow \citet{Shan2025}.
The primary mass follows a power-law plus Gaussian-peak model,
\begin{equation}
p(m_1)=
\left[(1-\lambda_{\rm peak})B(m_1)
+\lambda_{\rm peak}G(m_1)\right]S(m_1),
\end{equation}
where $B(m_1)\propto m_1^{-\alpha}$ is the power-law component, $G(m_1)$ is the Gaussian peak, and $S(m_1)$ is the low-mass smoothing function.
The secondary mass is sampled from
\begin{equation}
p(m_2|m_1)\propto
\left(\frac{m_2}{m_1}\right)^{\beta_q}S(m_2),
\qquad m_2<m_1 .
\end{equation}
The parameters used in the simulation are $\lambda_{\rm peak}=0.1$, $\alpha=2.63$, $m_{\rm min}=4.59\,M_\odot$, $m_{\rm max}=86.22\,M_\odot$, $\delta_m=4.82\,M_\odot$, $\mu_m=33.07\,M_\odot$, $\sigma_m=5.69\,M_\odot$, and $\beta_q=1.26$~\citep{2021ApJ...913L...7A}.
The aligned spin components $\chi_{1z}$ and $\chi_{2z}$ are sampled uniformly in $[0,0.99]$.
The inclination, sky position, polarization angle, and coalescence time are sampled from isotropic or uniform distributions.
\shan{More recent LVK population analyses indicate additional structure in the BBH mass distribution and provide updated constraints on the spin and redshift distributions \citep{2026arXiv260527226T}. Although adopting an updated population model may modify the quantitative weighting of events with different source parameters and SNR, we do not expect it to change our main qualitative conclusions, which are primarily determined by the cluster lens models and the local lensing environments.}
The unlensed BBH waveforms are generated using the \texttt{IMRPhenomXP} waveform approximant~\citep{Pratten_2021}, with the in-plane spin components set to zero.

\subsection{Strong-lensing optical depth and cluster lens macro model}
For a source at redshift $z_s$, the probability of being strongly lensed by a galaxy cluster is determined by the optical depth
\begin{equation}
\begin{split}
\tau(z_s) &=
\int_0^{z_s} dz_l
\int d\ln M_{200m}
\int dq\,
\frac{dV_c}{dz_l d\Omega}\,
\frac{dn}{d\ln M_{200m}}\,
p(q|z_l) \\
&\quad \times
S_{\rm cr}(M_{200m},z_l,z_s;q).
\end{split}
\end{equation}
Here, $z_l$ is the lens redshift, $dV_c/(dz_l\,d\Omega)$ is the 
comoving volume element per steradian, $dn/d\ln M_{200m}$ is the 
halo mass function from \citet{Tinker2008}, $q$ is the projected 
minor-to-major axis ratio of the smooth cluster halo, and 
$p(q|z_l)$ is its redshift-dependent distribution. 
\shan{The quantity $S_{\rm cr}$ is the source-plane strong-lensing cross 
section, defined as
$S_{\rm cr}\equiv\int_{\mathcal{R}_{\rm mult}}d^2\boldsymbol{\beta}$,
where $\mathcal{R}_{\rm mult}$ denotes the region of source positions 
for which the lens equation produces multiple macro images. 
We evaluate this area numerically from the caustic structure of the 
lens model and express it as a solid angle in steradians.}


The projected axis ratio $q$ follows the redshift-dependent $\beta$-distribution fit of \citet{Suto2016},
\begin{equation}
p(q|z_l)=
\frac{q^{a_q(z_l)-1}(1-q)^{b_q(z_l)-1}}
{B[a_q(z_l),b_q(z_l)]},
\end{equation}
where $B(a,b)$ is the Euler $\beta$ function.
The parameters $a_q$ and $b_q$ are linearly interpolated between the redshift nodes listed in Table~\ref{tab:suto_q_nodes}.

\begin{table}
\centering
\caption{Projected halo axis-ratio $\beta$-distribution parameters adopted from \protect\citet{Suto2016}.}
\label{tab:suto_q_nodes}
\begin{tabular}{ccc}
\toprule
$z_l$ & $a_q$ & $b_q$ \\
\midrule
0.0 & 4.35 & 3.39 \\
0.2 & 4.34 & 3.65 \\
0.4 & 4.21 & 3.69 \\
1.0 & 3.93 & 3.80 \\
\bottomrule
\end{tabular}
\end{table}

We use $M_{200m}$ as the lens mass because both the halo mass function and the adopted concentration relation are evaluated for a spherical overdensity halo mass defined with respect to the mean matter density.
In this definition, $M_{200m}$ is the mass enclosed by $r_{200m}$, where the mean density is 200 times the cosmic mean matter density:
\begin{equation}
M_{200m}
=
\frac{4\pi}{3}r_{200m}^3
\left[200\,\rho_m(z_l)\right],
\end{equation}
with $\rho_m(z_l)=\Omega_m\rho_{\rm crit,0}(1+z_l)^3$.

The halo mass range used here is
\begin{equation}
10^{14}M_\odot < M_{200m} < 10^{15.2}M_\odot .
\end{equation}
We focus on the cluster-scale mass range, rather than the full group- and cluster-scale range used in \citet{Chen2024}, because group-scale lenses lie in a transition regime between galaxy and cluster lenses.
In this regime, the macro lens model and the stellar microlensing model are less well defined, since the central galaxy, satellite galaxies, and diffuse stellar component can vary strongly from system to system.

We then use rejection sampling to decide whether a mock source is strongly lensed.
For each source, we draw a random number $u$ from a uniform distribution in $[0,1]$.
If $u<\tau(z_s)$, the source is selected as an event lensed by a cluster; otherwise, it is kept as an unlensed event.
In the mock survey, this selection is repeated until 300 detectable lensed events with network SNR $>8$ are obtained.
Here, we adopt the O4-sensitivity power spectral density (PSD) implemented as \texttt{aLIGOAdVO4T1800545} in PyCBC~\citep{Biwer:2018osg}, and use the LIGO--Virgo HLV detector network, i.e., H1, L1, and V1.
We do not re-estimate the lensing rate because \citet{Chen2024} have already estimated this rate.

For each accepted lensed event, we sample the lens redshift, halo mass, and projected axis ratio from the optical-depth-weighted lens population.
The target sampling distribution is
\begin{equation}
\label{eq:lens_kernal}
\begin{split}
p(z_l,\ln M_{200m},q|z_s)
&\propto
\frac{dV_c}{dz_l d\Omega}\,
\frac{dn}{d\ln M_{200m}}\,
p(q|z_l) \\
&\quad \times
S_{\rm cr}(M_{200m},z_l,z_s;q).
\end{split}
\end{equation}
This choice makes the accepted lens population consistent with the optical depth.

For each cluster, we model the smooth dark-matter halo as an elliptical NFW component and add explicit baryonic and substructure components.
The three-dimensional NFW profile is used to set the radial mass normalization \citep{NavarroFrenkWhite1997},
\begin{equation}
\rho_{\rm NFW}(r)=
\frac{\rho_s}
{(r/r_s)(1+r/r_s)^2},
\end{equation}
where $r_s$ is the scale radius and $\rho_s$ is the characteristic density.
The mass $M_{200m}$ is the lens mass sampled from the kernel above, so
\begin{equation}
r_{200m}
=
\left[
\frac{3M_{200m}}
{4\pi\,200\,\rho_m(z_l)}
\right]^{1/3}.
\end{equation}
The concentration is assigned using the relation in \citet{Duffy2008},
\begin{equation}
c_{200m} =
10.14
\left(
\frac{M_{200m}}{2\times10^{12}\,h^{-1}M_\odot}
\right)^{-0.081}
(1+z_l)^{-1.01},
\end{equation}
where $h=0.68$.
The scale radius is
\begin{equation}
r_s=\frac{r_{200m}}{c_{200m}}.
\end{equation}
Finally, the NFW normalization is fixed by
\begin{equation}
M_{200m}=4\pi\rho_s r_s^3
\left[
\ln(1+c_{200m})-\frac{c_{200m}}{1+c_{200m}}
\right].
\end{equation}
This equation sets $\rho_s$ and ensures that the NFW mass enclosed within $r_{200m}$ is equal to the sampled halo mass $M_{200m}$.
In the lensing calculation, this spherical NFW mass normalization is projected as an elliptical NFW lens model.
The ellipticity of the projected lens is specified by the sampled minor-to-major axis ratio $q$ and a position angle drawn uniformly in $[0,\pi)$.

For the BCG component, the stellar mass is assigned from the empirical BCG--cluster mass relation of \citet{2019A&A...631A.175E},
\begin{equation}
\label{eq:M200_bcg}
\log_{10}M_{\rm BCG}
=
t_1\log_{10}M_{200c}+t_2 ,
\end{equation}
where $M_{\rm BCG}$ and $M_{200c}$ are in units of $M_\odot$.
We use $(t_1,t_2)=(0.41,5.59)$ for $z_l\leq0.3$ and $(0.31,7.00)$ for $z_l>0.3$.

The BCG is modeled with the \textsc{lenstronomy} \texttt{HERNQUIST\_ELLIPSE\_CSE} lens profile~\citep{Hernquist1990,2021PASP..133g4504O}.
Here, CSE denotes the cored steep ellipsoid approximation, which represents the projected elliptical Hernquist profile as a superposition of analytic CSE components for fast lensing calculations~\citep{2021PASP..133g4504O}.
In \textsc{lenstronomy}, this profile is specified by the lensing normalization $\sigma_{0,\rm BCG}$, the angular scale radius $R_{s,\rm BCG}$, the ellipticity components $(e_{1,\rm BCG},e_{2,\rm BCG})$, and the lens center.

The physical Hernquist model is specified by the total stellar mass $M_{\rm BCG}$ and the Hernquist scale radius $a_{\rm BCG}$.
The projected effective radius of a Hernquist profile is related to this scale radius by
\begin{equation}
R_{e,\rm BCG}=1.8153\,a_{\rm BCG}.
\end{equation}
In the BCG+ICL stellar-mass fit described in Sec.~\ref{subsec:Microlensing}, we fit the projected annular stellar mass with a Hernquist profile and obtain $R_{e,\rm BCG}$.
The corresponding Hernquist scale radius $a_{\rm BCG}$ is then used to define the lensing parameters.
For a lens at redshift $z_l$ and a source at redshift $z_s$, the angular scale radius and normalization are
\begin{equation}
R_{s,\rm BCG}
=
\frac{a_{\rm BCG}}{D_l},
\end{equation}
and
\begin{equation}
\sigma_{0,\rm BCG}
=
\frac{M_{\rm BCG}}
{2\pi a_{\rm BCG}^2\Sigma_{\rm crit}(z_l,z_s)} ,
\end{equation}
where $D_l$ is the angular-diameter distance to the lens and $\Sigma_{\rm crit}$ is the critical surface density.

The projected shape is specified by the BCG axis ratio $q_{\rm BCG}$ and position angle $\phi_{\rm BCG}$.
They are converted to the \textsc{lenstronomy} ellipticity components as
\begin{equation}
\begin{split}
e_{1,\rm BCG}
&=
\frac{1-q_{\rm BCG}}{1+q_{\rm BCG}}
\cos 2\phi_{\rm BCG}, \\
e_{2,\rm BCG}
&=
\frac{1-q_{\rm BCG}}{1+q_{\rm BCG}}
\sin 2\phi_{\rm BCG}.
\end{split}
\end{equation}
We fix $q_{\rm BCG}=0.63$, motivated by the BCG shape measurements of \citet{Okabe2020}.
The BCG position angle is drawn relative to the smooth dark-matter halo position angle, with a mean absolute misalignment of $22.2^\circ$, following the same measurements.
The BCG center is fixed to the smooth cluster-halo center.

In addition to the dark-matter halo and the BCG, the macro lens model also includes the smooth stellar contribution from the ICL.
The ICL surface density is modeled as a circular power-law profile,
\begin{equation}
\label{eq:icl_power_law}
\Sigma_{\rm ICL}(R)
=
\Sigma_{50}
\left(\frac{R}{50\,{\rm kpc}}\right)^{-p}.
\end{equation}
The parameters $\Sigma_{50}$ and $p$ are obtained by fitting the projected ICL stellar masses in the outer radial annuli, as described in Sec.~\ref{subsec:Microlensing}.
The smooth macro model is therefore composed of an elliptical NFW halo, an elliptical Hernquist BCG, and an ICL component.

\begin{figure*}
\centering
\includegraphics[width=0.8\textwidth]{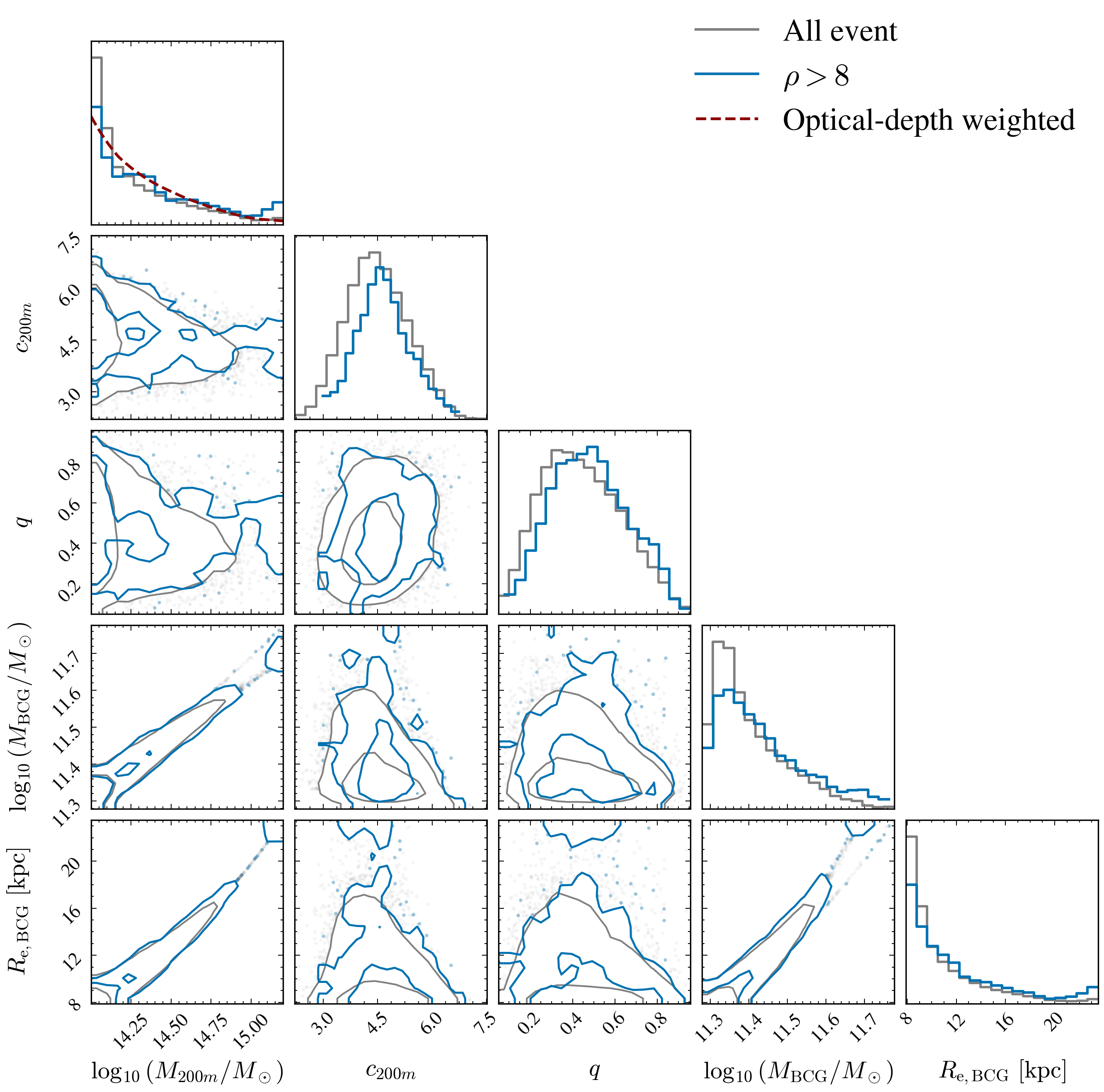}
\caption{Distribution of the cluster lenses used in the simulation. 
The corner plot shows the joint and marginalized distributions of $\log_{10}(M_{200m}/M_\odot)$, $c_{200m}$, $q$, $\log_{10}(M_{\rm BCG}/M_\odot)$, and $R_{\rm e,BCG}$. 
Grey points, contours, and histograms show all simulated lensed systems, while blue points, contours, and histograms show the subset with at least one macro image satisfying $\rho>8$. 
The dashed dark-red curve in the $\log_{10}(M_{200m}/M_\odot)$ panel shows the optical-depth-weighted lens-mass distribution, i.e., the $M_{200m}$ marginal distribution of the lens-population kernel in Eq.~(\ref{eq:lens_kernal}), averaged over the simulated source-redshift distribution.
}
\label{fig:lens_mass_bcg_mass}
\end{figure*}

Figure~\ref{fig:lens_mass_bcg_mass} shows the parameter distribution of the cluster lenses used in the simulation.
The corner plot gives the joint and marginalized distributions of $\log_{10}(M_{200m}/M_\odot)$, $c_{200m}$, $q$, $\log_{10}(M_{\rm BCG}/M_\odot)$, and $R_{\rm e,BCG}$.
The grey points, contours, and histograms show all simulated lensed systems, while the blue points, contours, and histograms show the subset with at least one macro image satisfying $\rho>8$.
The dashed dark-red curve in the $\log_{10}(M_{200m}/M_\odot)$ panel shows the optical-depth-weighted lens-mass distribution, namely the $M_{200m}$ marginal distribution of the lens-population kernel in Eq.~(\ref{eq:lens_kernal}), averaged over the simulated source-redshift distribution.
One can find that the sampled lenses are concentrated toward the lower end of the adopted cluster-mass interval because the halo number density decreases rapidly with mass.

\subsection{Satellite galaxies and substructures}
\label{subsec:SatelliteGalaxies}

For each cluster realization, we add satellite galaxies and their associated dark-matter subhalos on top of the smooth macro model.
The accreted subhalo population is generated following the prescription of \citet{Han2016}, in which the accreted subhalo masses are sampled from
\begin{equation}
\frac{dN}{dm_{\rm acc}}
\propto
m_{\rm acc}^{-(1+\alpha_{\rm sub})},
\end{equation}
with $\alpha_{\rm sub}=0.95$.
The expected number of accreted systems is
\begin{equation}
\begin{split}
\langle N_{\rm sub}\rangle
&=
A_{\rm acc}
\frac{M_{200c}}{M_0} \\
&\quad \times
\frac{
(m_{\rm acc,min}/M_0)^{-\alpha_{\rm sub}}
-
(m_{\rm acc,max}/M_0)^{-\alpha_{\rm sub}}
}
{\alpha_{\rm sub}},
\end{split}
\end{equation}
where $M_{200c}$ is the host halo mass enclosed within $r_{200c}$, defined by a mean interior density of $200\rho_{\rm crit}(z_l)$, $M_0=10^{10}h^{-1}M_\odot$, and $A_{\rm acc}=0.080$.
The actual number of accreted systems is then drawn as
\begin{equation}
N_{\rm sub}\sim {\rm Poisson}\left(\langle N_{\rm sub}\rangle\right),
\end{equation}
so that different cluster realizations have integer subhalo counts whose ensemble mean is $\langle N_{\rm sub}\rangle$.
After $N_{\rm sub}$ is drawn, the individual accreted masses are sampled from the power-law mass function above over the range $[m_{\rm acc,min},m_{\rm acc,max}]$.
To balance computational cost and completeness for the massive satellite galaxies that dominate the local stellar surface density, we adopt
\begin{equation}
\begin{split}
&m_{\rm acc,min}=10^{11}M_\odot,
\\
&m_{\rm acc,max}=\min(0.01M_{200c},10^{13}M_\odot).
\end{split}
\end{equation}

The spatial positions of the accreted systems are sampled from an NFW radial number-density profile with the same $r_{200c}$ and $c_{200c}$ as the host halo, and their angular directions are drawn isotropically.
We then apply the survival and mass-loss model of \citet{Han2016}.
A system is retained with survival probability $f_{\rm surv}=0.56$.
For a surviving system at three-dimensional radius $r$, the bound-to-accreted mass ratio is sampled as
\begin{equation}
\begin{split}
\ln\frac{m_{\rm bound}}{m_{\rm acc}}
&\sim
\mathcal{N}
\left[
\ln\left(
0.34\frac{r}{r_{200c}}
\right),
1.1^2
\right].
\end{split}
\end{equation}
This step converts the accreted mass $m_{\rm acc}$ into the present-day bound mass $m_{\rm bound}$, accounting for tidal mass loss after accretion.
We keep only systems with projected radius smaller than $\min(500\,{\rm kpc},r_{200c})$ and bound mass $m_{\rm bound}>10^9M_\odot$.
This cut restricts the explicit substructure realization to systems that can contribute appreciably to the lensing and local stellar surface density, while avoiding a large number of low-mass or distant subhalos with negligible effects but high computational cost.

The dark-matter component of each retained system is modeled as a truncated NFW subhalo using the Baltz--Marshall--Oguri truncation form~\citep{Baltz2009}.
The infall NFW profile is specified by $m_{\rm acc}$ and its concentration, which is assigned using the \citet{Duffy2008} concentration--mass relation.
The truncation radius is then chosen such that the total truncated mass is equal to the sampled bound mass $m_{\rm bound}$.

Each retained system is also assigned a stellar satellite-galaxy component.
The stellar mass is obtained from the stellar-to-halo mass relation of \citet{Moster2013},
\begin{equation}
\frac{M_\star}{m_{\rm acc}}
=
2N(z_l)
\left[
\left(\frac{m_{\rm acc}}{M_1(z_l)}\right)^{-\beta(z_l)}
+
\left(\frac{m_{\rm acc}}{M_1(z_l)}\right)^{\gamma(z_l)}
\right]^{-1}.
\end{equation}
The redshift-dependent parameters follow \citet{Moster2013}:
\begin{equation}
\log_{10}M_1(z_l)
=
M_{10}+M_{11}\frac{z_l}{1+z_l},
\end{equation}
\begin{equation}
N(z_l)
=
N_{10}+N_{11}\frac{z_l}{1+z_l},
\end{equation}
\begin{equation}
\begin{split}
\beta(z_l)
&=
\beta_{10}+\beta_{11}\frac{z_l}{1+z_l}, \\
\gamma(z_l)
&=
\gamma_{10}+\gamma_{11}\frac{z_l}{1+z_l}.
\end{split}
\end{equation}
We adopt
\begin{equation}
\begin{split}
(M_{10},M_{11})&=(11.590,1.195), \\
(N_{10},N_{11})&=(0.0351,-0.0247),
\end{split}
\end{equation}
and
\begin{equation}
\begin{split}
(\beta_{10},\beta_{11})&=(1.376,-0.826), \\
(\gamma_{10},\gamma_{11})&=(0.608,0.329).
\end{split}
\end{equation}
We include a log-normal scatter of $0.15$ dex and impose $M_\star\leq0.5m_{\rm acc}$.

The effective radius is assigned using the early-type galaxy size--mass relation of \citet{Shen2003},
\begin{equation}
R_e=2.88\times10^{-6}
\left(\frac{M_\star}{M_\odot}\right)^{0.56}
{\rm kpc}.
\end{equation}
The stellar component is represented by a circular Hernquist lens profile with scale radius $a=R_e/1.8153$.
Because the adopted accreted-mass range starts at $10^{11}M_\odot$, we do not impose an additional stochastic galaxy-occupation model; all retained systems are assigned stellar components through the stellar-to-halo mass relation above.

When the explicit stellar and subhalo components are added, the smooth NFW normalization is reduced to keep the total cluster mass budget fixed.
In the pipeline, the sampled $M_{200m}$ halo is first converted to the corresponding $M_{200c}$ NFW halo.
The smooth-halo deflection normalization is then multiplied by
\begin{equation}
\begin{split}
f_{\rm sm}
&=
1-
\frac{
M_{\rm BCG}
+M_{\rm ICL}(<r_{200c})
+M_{\rm sub,bound}
+M_{\rm sat,*}
}{
M_{200c}
},
\end{split}
\end{equation}
where $M_{\rm sub,bound}=\sum_i m_{{\rm bound},i}$ is the total bound mass of the retained dark-matter subhalos, $M_{\rm sat,*}=\sum_i M_{\star,i}$ is the total stellar mass of the retained satellite galaxies, and $M_{\rm ICL}(<r_{200c})$ is the ICL mass enclosed within $r_{200c}$.

\subsection{Microlensing effect on the macro images}
\label{subsec:Microlensing}

For each accepted source, we randomly sample a source position from the numerically constructed caustic-envelope region and solve the lens equation
\begin{equation}
\boldsymbol{y}
=
\boldsymbol{x}
-
\boldsymbol{\alpha}(\boldsymbol{x}),
\end{equation}
where $\boldsymbol{x}$ and $\boldsymbol{y}$ are the image-plane and source-plane angular positions, and $\boldsymbol{\alpha}$ is the total deflection angle of the composite cluster lens.
We retain systems with at least two macro images.
At each image position, we compute the macro convergence $\kappa$ and shear $\gamma$.

The microlensing strength is determined by the stellar convergence at each macro-image position,
\begin{equation}
\kappa_\star
=
\frac{\Sigma_\star}{\Sigma_{\rm crit}},
\end{equation}
where $\Sigma_\star$ is the projected stellar surface density at the image position and $\Sigma_{\rm crit}$ is the lensing critical surface density.
The projected physical distance of the image from the cluster center is
\begin{equation}
R =
D_l
\sqrt{
\theta_{1,\rm img}^2+
\theta_{2,\rm img}^2
},
\end{equation}
\shan{where $D_l$ is the angular-diameter distance to the lens, and
$\theta_{1,\rm img}$ and $\theta_{2,\rm img}$ are the angular coordinates
of the image relative to the cluster center, expressed in radians.}

We calibrate the BCG+ICL stellar surface-density model using the annular BCG+ICL stellar-mass measurements from the Dark Energy Survey (DES) and Atacama Cosmology Telescope (ACT) cluster sample of \citet{Golden_Marx_2023}.
For annulus $a$, the mean stellar mass is written as
\begin{equation}
\label{eq:annular_stellar_mass}
\begin{split}
\log_{10}\left(\frac{M_{\star,\rm ann,a}}{h^{-2}M_\odot}\right)
&=
11.5+\alpha_{{\rm ann},a} \\
& +
\beta_{{\rm ann},a}
\left[
\log_{10}\left(\frac{M_{200m}}{M_\odot}\right)-14.80
\right] \\
&-0.25 .
\end{split}
\end{equation}
The term $-0.25$ converts the stellar masses from a Salpeter initial mass function (IMF) to a Chabrier IMF.
We use only the mean relation in Eq.~(\ref{eq:annular_stellar_mass}) and do not draw stochastic scatter realizations.
The intrinsic scatter $\sigma_{\rm int}$ is used only to normalize the logarithmic residuals in the ICL and BCG profile fits below.
The annuli and coefficients used in these fits are summarized in Table~\ref{tab:icl_annular_mass}.

\begin{table}
\centering
\caption{Coefficients of the annular BCG+ICL stellar mass relation. The first column gives the projected radial annulus. The parameters $\alpha_{\rm ann}$ and $\beta_{\rm ann}$ define the halo-mass dependence of $M_{\star,\rm ann}$ in Eq.~(\ref{eq:annular_stellar_mass}), and $\sigma_{\rm int}$ is the intrinsic scatter in stellar mass at fixed halo mass. These values are taken from the no-core BCG+ICL measurements of \citet{Golden_Marx_2023}.}
\label{tab:icl_annular_mass}
\begin{tabular}{cccc}
\toprule
Annulus (kpc) & $\alpha_{\rm ann}$ & $\beta_{\rm ann}$ & $\sigma_{\rm int}$ \\
\midrule
$0$ to $10$ & -0.059 & 0.051 & 0.133 \\
$10$ to $30$ & 0.023 & 0.235 & 0.186 \\
$30$ to $50$ & -0.366 & 0.463 & 0.268 \\
$50$ to $100$ & -0.206 & 0.297 & 0.256 \\
$50$ to $150$ & 0.101 & 0.442 & 0.251 \\
$100$ to $300$ & 0.417 & 0.384 & 0.289 \\
$50$ to $300$ & 0.507 & 0.372 & 0.285 \\
$150$ to $300$ & 0.317 & 0.489 & 0.296 \\
\bottomrule
\end{tabular}
\end{table}

We first determine the smooth ICL profile from the outer annuli.
For the circular power-law profile in Eq.~(\ref{eq:icl_power_law}), the predicted ICL mass in annulus $a$ is
\begin{equation}
M_{{\rm ICL},a}
=
2\pi
\int_{R_{{\rm in},a}}^{R_{{\rm out},a}}
R\,\Sigma_{50}
\left(\frac{R}{50\,{\rm kpc}}\right)^{-p}
dR .
\end{equation}
The parameters $\Sigma_{50}$ and $p$ are obtained by fitting the annular masses in the $50$--$100\,{\rm kpc}$, $50$--$150\,{\rm kpc}$, $100$--$300\,{\rm kpc}$, $50$--$300\,{\rm kpc}$, and $150$--$300\,{\rm kpc}$ annuli.
Specifically, we minimize the scatter-normalized residuals
\begin{equation}
\chi^2_{\rm ICL}
=
\sum_a
\left[
\frac{
\log_{10}M_{{\rm ICL},a}
-
\log_{10}M_{\star,\rm ann,a}
}
{\sigma_{{\rm int},a}}
\right]^2 ,
\end{equation}
where the sum is over the outer annuli listed above.
The slope is restricted to $0.01<p<1.95$.

After the ICL profile is fixed, we determine the BCG scale radius from the inner residual stellar mass.
For the $0$--$10\,{\rm kpc}$, $10$--$30\,{\rm kpc}$, and $30$--$50\,{\rm kpc}$ annuli, we subtract the ICL contribution predicted by the fitted power law:
\begin{equation}
M_{{\rm BCG,res},a}
=
M_{\star,\rm ann,a}
-
M_{{\rm ICL},a}.
\end{equation}
We then fit a projected Hernquist profile to these residual annular masses.
The fitted Hernquist scale radius is reported through the equivalent circularized effective radius,
\begin{equation}
R_{e,\rm BCG}=1.8153\,a_{\rm BCG}.
\end{equation}
This is the same scale-radius convention used for the BCG lens component in the macro model.
In the final lens model, we use this fitted scale radius $a_{\rm BCG}$, but normalize the total BCG stellar mass to the empirical $M_{\rm BCG}$--$M_{200c}$ relation in Eq.~(\ref{eq:M200_bcg}).

\shan{The stellar convergence at each macro-image position is computed from
the combined projected stellar surface density of the BCG, ICL, and
satellite galaxies:
\begin{equation}
\kappa_\star
=
\frac{
\Sigma_{\rm BCG}
+
\Sigma_{\rm ICL}
+
\Sigma_{\rm sat}
}{
\Sigma_{\rm crit}
}.
\end{equation}
Here, $\Sigma_{\rm BCG}$ is evaluated from the elliptical Hernquist
BCG component, $\Sigma_{\rm ICL}$ from
Eq.~(\ref{eq:icl_power_law}), and $\Sigma_{\rm sat}$ from the circular
Hernquist profiles of the satellite galaxies described in
Sec.~\ref{subsec:SatelliteGalaxies}.}

\begin{figure*}
\centering
\includegraphics[width=\textwidth]{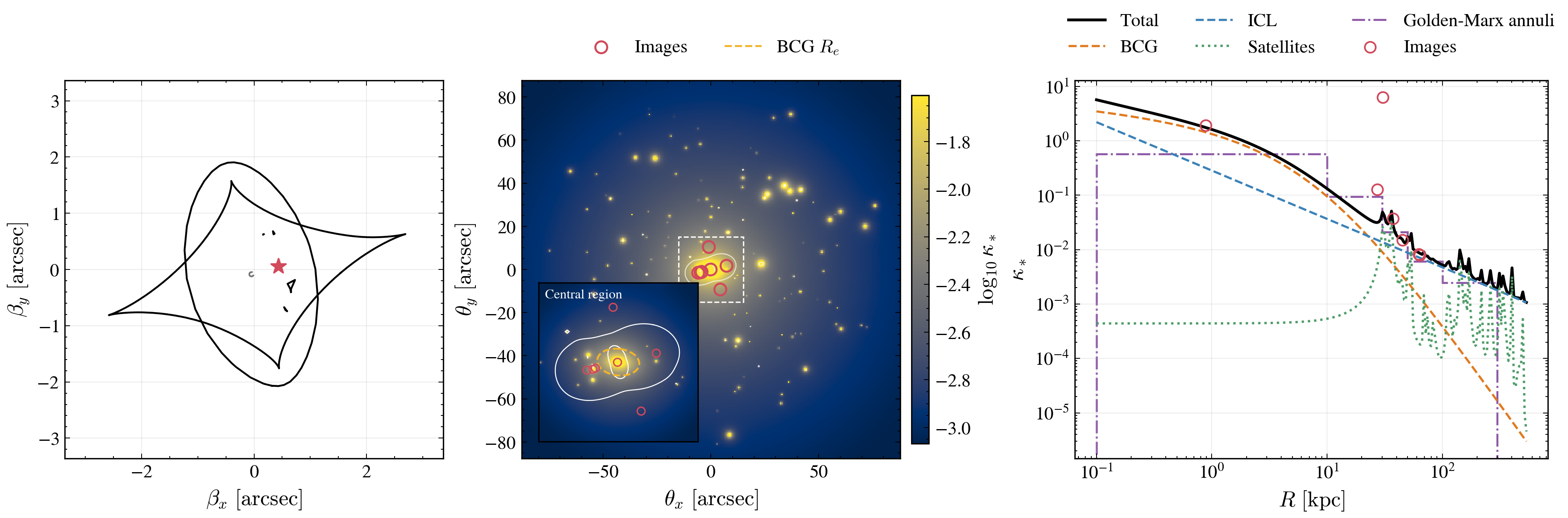}
\caption{Example of a simulated cluster-lensing system and its local stellar convergence. The left panel shows the source plane, with the source position marked by the red star and the caustic structure shown by the black curves. The middle panel shows the image plane. The colour map gives $\log_{10}\kappa_\star$, and open red circles mark the lensed images. The inset zooms into the central region around the macro images; the yellow dashed ellipse marks the projected BCG effective radius. The right panel shows the radial stellar-convergence profile. The black, orange dashed, blue dashed, and green dotted curves show the \shan{azimuthally averaged} total, BCG, ICL, and satellite contributions, respectively. The purple dash-dotted steps show the annular stellar convergence inferred from the BCG+ICL stellar-mass relation of \citet{Golden_Marx_2023}, and the open red circles mark the image positions.
}
\label{fig:example_lens_source_image_stellar}
\end{figure*}

Figure~\ref{fig:example_lens_source_image_stellar} shows a representative
cluster-lensing realization.
The left panel shows the source plane, where the red star marks the source
position and the black curves show the caustics of the composite cluster
lens model.
The main caustic structure is produced by the smooth cluster halo and the
BCG, while the smaller caustics are associated with satellite galaxies.
The middle panel shows the image plane.
The open red circles mark the macro-image positions, the white curve shows
the critical curve, and the colour scale gives $\log_{10}\kappa_\star$
from the combined BCG, ICL, and satellite-galaxy components.
The right panel shows the azimuthally averaged radial stellar-convergence
profile.
The black solid, orange dashed, and blue dashed curves show the total,
BCG, and ICL contributions, respectively, while the green dotted curves
show the individual satellite-galaxy contributions.
The purple dash-dotted steps show the annular stellar convergence inferred
from the annular stellar-mass relation of
\citet{Golden_Marx_2023}.
The open red circles indicate the local values of $\kappa_\star$ at the
macro-image positions.
Because the curves are azimuthally averaged whereas the satellite-galaxy
contribution is spatially localized, the image-position values need not
lie on the total radial profile.

One can find that the total profile broadly follows the annular constraints, and that the inner region is dominated by the BCG.
The ICL contribution becomes important at radii of the order of several tens of kpc.
Although the satellite-galaxy contribution is smaller than the ICL contribution in the azimuthally averaged profile, satellite galaxies are more spatially concentrated.
Therefore, if a macro image happens to be close to a satellite galaxy, its local $\kappa_\star$ can be dominated by the satellite-galaxy component, as shown by the point with the highest $\kappa_\star$ in the figure.
Thus, the composite BCG+ICL+satellite-galaxy model is needed to describe the microlensing density in cluster lenses.
Without the ICL and satellite-galaxy components, the stellar density would be underestimated at large radii.

After obtaining the macro convergence $\kappa$, shear $\gamma$, and stellar convergence $\kappa_\star$, we calculate the microlensing effect.
The compact population contains living stars sampled from a Chabrier IMF~\citep{Chabrier2003}, restricted to the mass range $0.08 M_\odot \leq m_\star \leq 1.5 M_\odot$.
The remnant population is generated based on the initial-to-final mass relation of \citet{Spera2015}.
In the current simulation, the maximum remnant mass is set to $28\,M_\odot$.
The ratio of remnant mass to living-star mass is fixed to $0.48$, consistent with the Chabrier-IMF stellar remnant budget in~\citet{MadauDickinson2014}.

Following the standard wave-optics treatment~\citep{TakahashiNakamura2003,1999PThPS.133..137N,Shan2025,Tambalo:2022plm,Villarrubia-Rojo:2024xcj}, the amplification factor is computed from the Fresnel--Kirchhoff diffraction integral,
\begin{equation}
F(\omega)
=
\frac{2G\langle M\rangle (1+z_l)\omega}
{\pi c^3 i}
\int d^2\boldsymbol{x}\,
\exp\left[i\omega t(\boldsymbol{x})\right],
\end{equation}
where $\omega=2\pi f$ is the angular frequency, $\langle M\rangle$ is the mean microlens mass in solar units, and $t(\boldsymbol{x})$ is the physical time-delay surface.
In the local coordinates centred on the macro image, the dimensionless part of the time-delay surface is
\begin{equation}
\begin{split}
\tau(\boldsymbol{x})
&=
\frac{1}{2}(1-\kappa+\gamma)x_1^2
+
\frac{1}{2}(1-\kappa-\gamma)x_2^2 \\
&\quad
-
\sum_i
\frac{m_i}{\langle M\rangle}
\ln|\boldsymbol{x}-\boldsymbol{x}_i| \\
&\quad
+
\tau_{\rm sheet}(\boldsymbol{x}),
\end{split}
\end{equation}
and the physical delay is
\begin{equation}
t(\boldsymbol{x})
=
\frac{4G\langle M\rangle}{c^3}
(1+z_l)\,
\tau(\boldsymbol{x}) .
\end{equation}
Here $m_i$ and $\boldsymbol{x}_i$ are the mass and position of the $i$th microlens, and $\tau_{\rm sheet}$ is the compensating smooth sheet term that keeps the total convergence fixed.
The numerical construction of the time-delay map and the Fourier transform of the area function follow the algorithms developed in \citet{Shan:2022xfx,Shan2025}; we do not repeat the details of the numerical grid and convergence tests here.

Finally, the microlensed GW waveform is obtained by multiplying the
unlensed waveform by the full amplification factor,
\begin{equation}
\tilde h_{\rm micro}(f)
=
F(f)\,\tilde h_{\rm unlensed}(f).
\end{equation}
The resulting waveforms are used to quantify microlensing-induced
distortions through the mismatch
\begin{equation}
{\cal M}(h_1,h_2)
=
1-
\max_{\Delta t,\phi}
\frac{(h_1|h_2)}
{\sqrt{(h_1|h_1)(h_2|h_2)}} ,
\label{eq:mismatch}
\end{equation}
where the maximisation is performed over a relative time shift
$\Delta t$ and a constant phase shift $\phi$. The noise-weighted
inner product is defined as
\begin{equation}
(a|b)
=
4\,{\rm Re}
\int_{f_{\rm low}}^{\infty}
\frac{\tilde a^*(f)\tilde b(f)}
{S_n(f)}
\,df .
\label{eq:inner_product}
\end{equation}

\shan{We consider two types of mismatch. The single-image mismatch compares
the microlensed waveform of each image with its corresponding
smooth macro-lensed waveform, thereby quantifying the waveform distortion
induced by microlensing for that image. The cross-mismatch compares the
microlensed waveforms of two detectable images of the same source,
thereby quantifying the difference between the microlensing distortions
experienced by the two images.}


\section{Results}
\label{sec:results}

In this section, we show the main properties of GWs lensed by clusters.
We focus on two aspects of cluster lensing: the time delays and magnification-ratios between detected image pairs, and the microlensing-induced waveform mismatch.

\subsection{Time-delay and magnification-ratio}
We first study the time-delay and magnification-ratio distributions of all simulated and detected lensed GW image pairs.
In this work, a macro image is treated as detectable if its network SNR is larger than 8.

\begin{figure*}
    \centering
    \includegraphics[width=\textwidth]{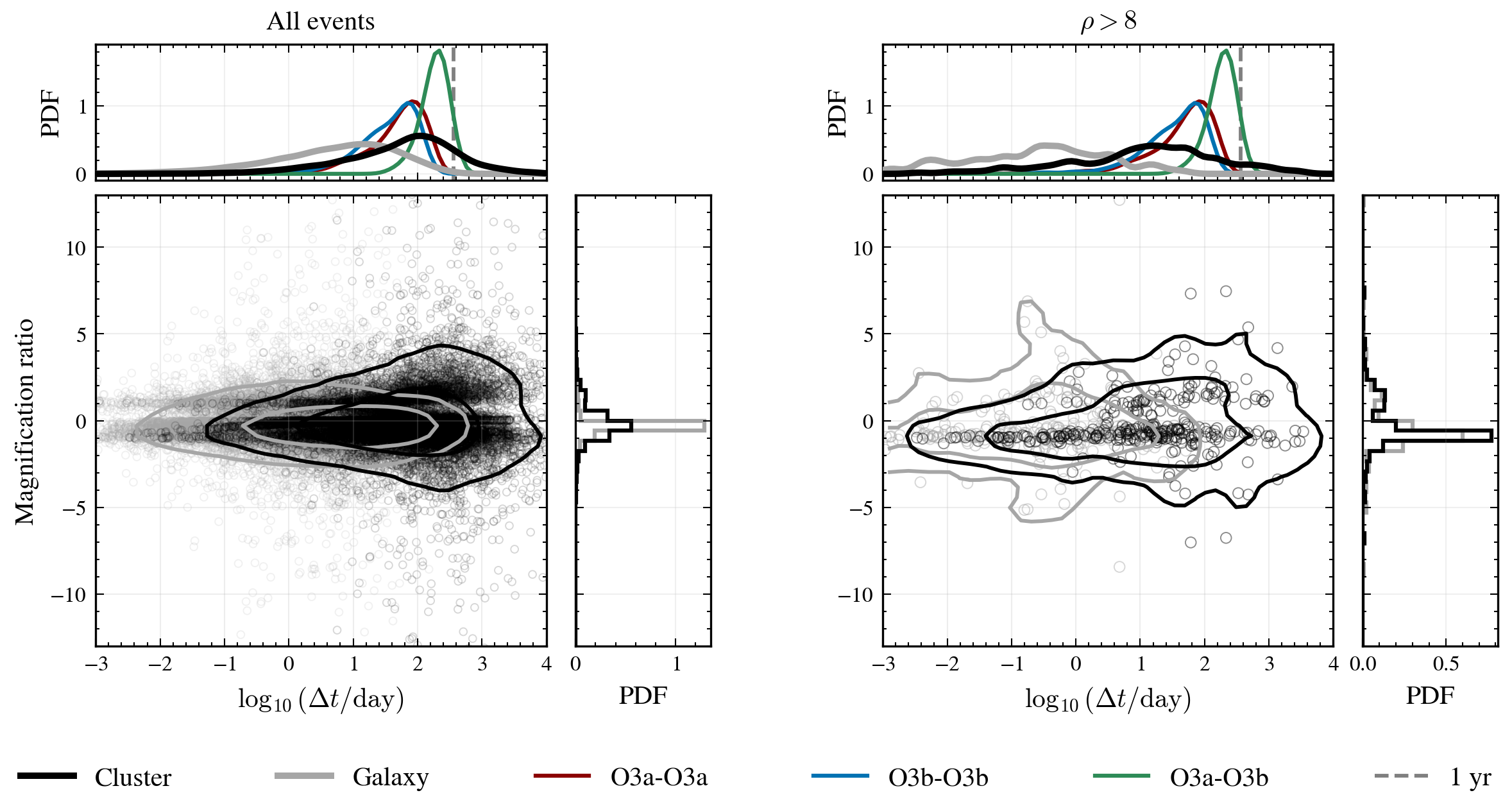}
    \caption{
    \shan{Time-delay and magnification-ratio distributions of lensed GW image pairs.} The left panel shows all simulated image pairs, while the right panel shows image pairs from systems with at least two macro images having network SNR $\rho>8$. The grey and black distributions correspond to the galaxy-lensing and cluster-lensing samples, respectively. The upper subpanel in each panel shows the marginalised time-delay distribution (normalized probability distribution function), while the coloured curves show the time separations of observed event pairs within O3a, within O3b, and between O3a and O3b.
    }
    \label{fig:delta_t_mu_cluster_galaxy_o4}
\end{figure*}

The results are shown in Figure~\ref{fig:delta_t_mu_cluster_galaxy_o4}. The left panel shows all simulated image pairs, while the right panel shows image pairs from systems with at least two macro images having network SNR $\rho>8$. The grey distribution represents the galaxy-lensing sample obtained using the galaxy-lensing optical depth and lens model described in \citet{Haris2018}, with the same source population as that described in Section~\ref{sec:methods}, while the black distribution represents the cluster-lensing sample. The coloured curves in the upper subpanel of each panel show the pair-separation distributions of public GWOSC O3 BBH events~\citep{KAGRA:2021vkt, Abbott_2023}, divided into O3a--O3a, O3b--O3b, and O3a--O3b pairs.

We find that the cluster-lensing time delays are much longer than the galaxy-lensing time delays.
In the all-event panel, the cluster-lensing probability density function (PDF) peaks at $\Delta t\simeq 110$ days, while the galaxy-lensing PDF peaks at $\Delta t\simeq 19$ days.
The corresponding median time delays are about $76$ days for cluster lenses and $6.9$ days for galaxy lenses.
After applying the SNR cut, the typical time delays of both the cluster and galaxy samples decrease significantly.
This is because the SNR cut removes many high-redshift sources and, in the cluster case, image pairs near the cluster centre, where the Shapiro time delay is very long.
However, the detected cluster sample still has much longer time delays than the detected galaxy sample: the median delay is about $15.4$ days for cluster lenses and $0.24$ days for galaxy lenses.

In addition, the detected cluster sample still retains a long-delay tail: about $9.1\%$ of image pairs have $\Delta t>365$ days, and about $3.3\%$ have $\Delta t>1000$ days.
In contrast, none of the detected galaxy-lensing pairs in the current sample reaches such long delays.
This tail is mainly due to the larger masses of cluster lenses.
At the same time, even for cluster lenses with masses larger than $10^{14}\,M_\odot$, some image pairs can still have very short time delays, down to about $0.01$ days.
These short-delay pairs correspond to close image pairs, whose number can be enhanced by substructure lensing.
This is illustrated by the three closely separated images in the middle panel of Figure~\ref{fig:example_lens_source_image_stellar}, and is qualitatively consistent with previous studies showing that cluster substructure or embedded subhalos can generate short-time-delay image pairs and enhance image multiplicities~\citep{Vujeva2025,Vujeva:2025nwg}.

From the comparison with the O3 pair-separation curves, the detected cluster-lensing distribution overlaps the observed within-run O3 separations much more strongly than the detected galaxy-lensing distribution.
The detected galaxy-lensing distribution is concentrated at sub-day to few-day delays after the SNR cut.
Therefore, a galaxy-lensing time-delay prior would still penalize many within-run event pairs.
If we consider O3a--O3b event pairs, the galaxy-lensing time-delay prior would penalize almost all pairs, while cluster lenses can still contribute.
These differences motivate the use of either a cluster-specific time-delay model or a complete population model that jointly accounts for galaxy-, group-, and cluster-scale lenses in searches for strongly lensed GW pairs. However, because the time-delay prior imposes a strong constraint on the selection of strongly lensed pairs, it is essential to adopt models that are as realistic as possible, including both the source-population and lensing models.

\shan{From the magnification-ratio distributions, one can see that, when all simulated image pairs are considered, the cluster-lensing sample has a broader distribution than the galaxy-lensing sample because of the inclusion of member galaxies in the cluster-lensing simulation, while the galaxy-lensing sample is more concentrated. However, for the detectable image pairs shown in the right panel, the two distributions become very similar. This suggests that, at O4 sensitivity, differences in the magnification-ratio distributions of different lens populations are unlikely to substantially bias the selection of strongly lensed GW candidates.}


\subsection{Microlensing properties}

\begin{figure}
    \centering
    \includegraphics[width=\linewidth]{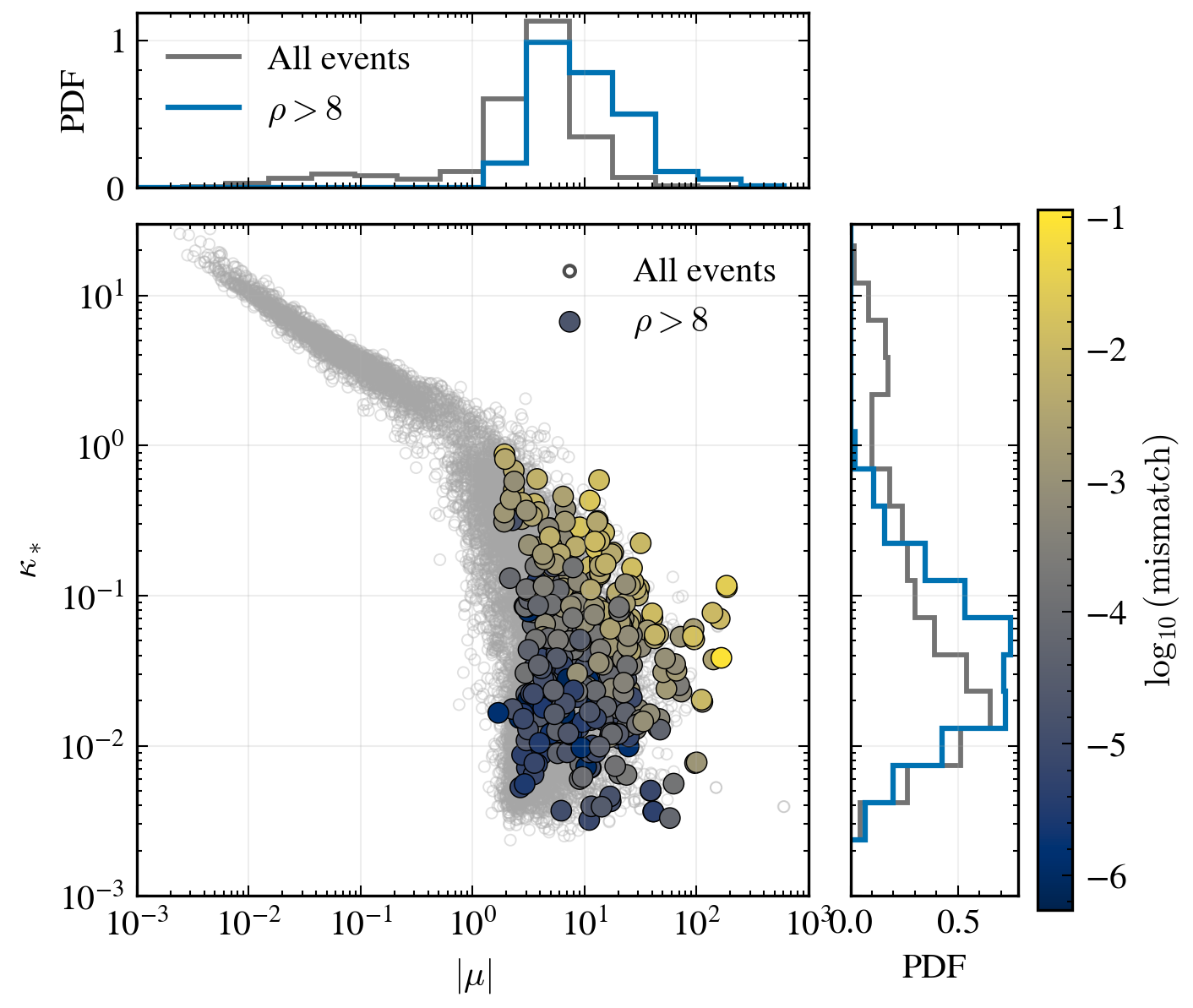}
    \caption{Relation between macro-image magnification, local stellar convergence, and microlensing mismatch for the cluster-lensing sample. Grey open circles show all simulated macro images, while filled circles show the subset with network SNR $\rho>8$. The colour of each filled circle gives the microlensing-induced mismatch $\log_{10}({\rm mismatch})$ \shan{between the microlensed waveform and the corresponding smooth macro-lensed waveform without microlensing.} The upper and right panels show the marginalised distributions of $|\mu|$ and $\kappa_\star$, respectively, for all macro images and for the $\rho>8$ subset.}
    \label{fig:magnification_kappa_star_mismatch}
\end{figure}

We next study the microlensing effects on GWs lensed by clusters.
Figure~\ref{fig:magnification_kappa_star_mismatch} shows the relation
between the absolute macro magnification $|\mu|$, the local stellar
convergence $\kappa_\star$, and the single-image microlensing mismatch
${\cal M}_{\rm single}$ for the cluster-lensing sample.
The marginal distributions show that the detected cluster images occupy
a narrower range of $\kappa_\star$ than the full image sample.
The detected images are also preferentially concentrated at higher
magnifications, with $|\mu|$ ranging from several to tens, owing to the
SNR selection.

In addition, the population-level result shows that not all cluster-lensed GW images have very low microlens densities.
Such low-density cases are often expected for highly magnified stars near the critical curves of cluster lenses~\citep{Venumadhav_2017}.
These critical curves are usually far from the BCG, so the local stellar density mainly comes from the ICL and satellite galaxies.
However, our GW image sample is a full strong-lensing sample, and therefore it also contains many images with relatively high stellar densities.
These high densities may occur because the images are close to the BCG, or because they are close to satellite galaxies, as shown in the right panel of Figure~\ref{fig:example_lens_source_image_stellar}.


One can see from Figure~\ref{fig:magnification_kappa_star_mismatch} that, for the detected images, the largest mismatches between each microlensed waveform and the corresponding smooth macro-lensed waveform without microlensing mainly occur for images with both high $\kappa_\star$ and high magnification.
This is because a higher macro magnification increases the effective microlens mass scale, while a higher stellar surface density gives a higher probability of sampling massive microlenses.

\begin{figure*}
    \centering
    \includegraphics[width=\textwidth]{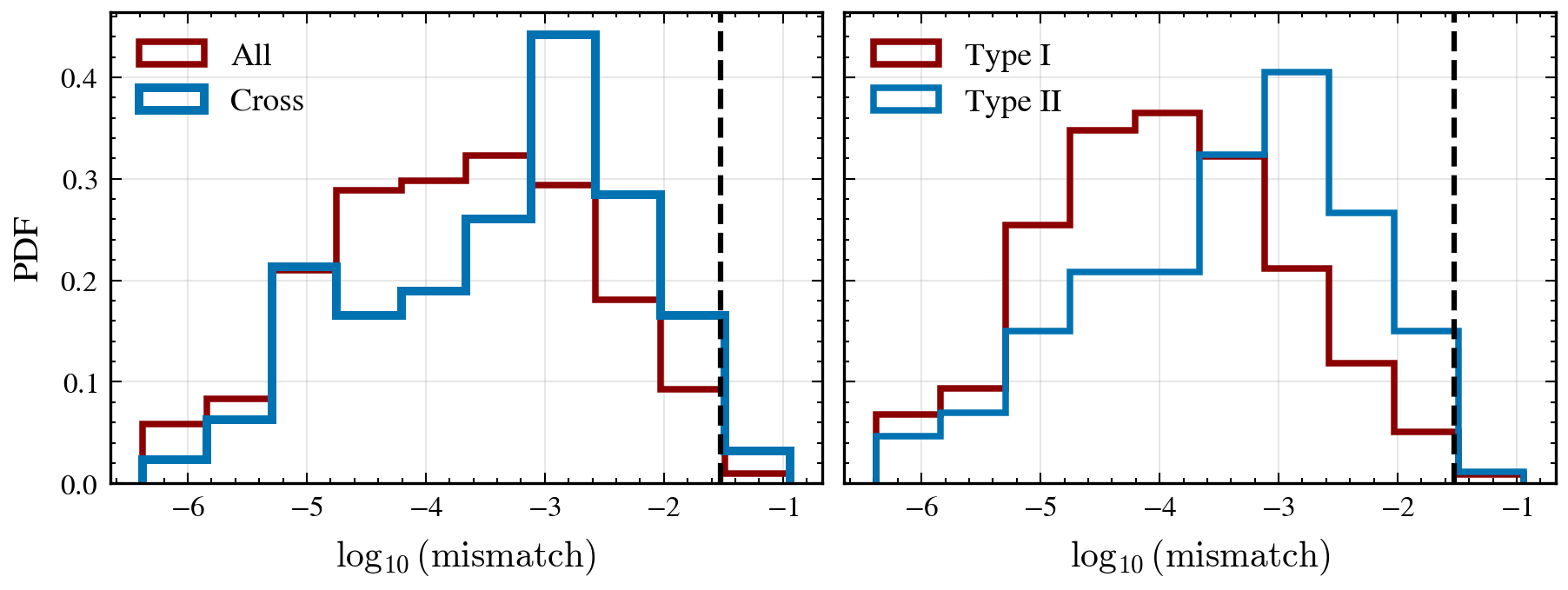}
    \caption{Waveform mismatch distributions induced by microlensing. The left panel shows the mismatch between each microlensed waveform and the corresponding smooth macro-lensed waveform, together with the cross-mismatch between microlensed waveforms from different macro images of the same source. \shan{The right panel shows the mismatch distributions between the microlensed and corresponding smooth macro-lensed waveforms for Type I and Type II macro images. The vertical dashed line marks ${\cal M}=0.03$. }
    }
    \label{fig:mismatch_histogram}
\end{figure*}

\shan{Figure~\ref{fig:mismatch_histogram} shows the mismatch distributions.
In the left panel, the red histogram shows the single-image mismatch
between each microlensed waveform and the corresponding smooth
macro-lensed waveform, while the blue histogram shows the cross-mismatch
between the microlensed waveforms of different detectable macro images
of the same source.
One can find that both distributions are concentrated well below
${\cal M}=0.03$, which is commonly adopted as a benchmark for waveform
accuracy in matched-filter searches~\citep{2009PhRvD..79l2001A,
2009PhRvD..80d7101A}.
Under the simplified assumption that the mismatch only reduces the
recovered matched-filter SNR, ${\cal M}=0.03$ corresponds to a loss of
less than $10\%$ in sensitive volume.
However, it should not be interpreted as a sharp detectability threshold,
because the recovery of distorted signals also depends on template-bank
optimisation, signal-consistency tests, and the ranking statistic used by
the search pipeline~\citep{Chan_2025}.
Only about $0.8\%$ of the single-image mismatches exceed
${\cal M}=0.03$, indicating that microlensing is unlikely to cause a
substantial loss of matched-filter SNR for most cluster-lensed images.
Similarly, only about $1.7\%$ of the cross-mismatches exceed this value,
showing that microlensing rarely produces large waveform differences
between detectable macro images of the same source.}

The right panel of Figure~\ref{fig:mismatch_histogram} separates the single-image mismatch by macro-image type.
We find that Type I images have smaller microlensing mismatches than Type II images.
The peak of the Type I distribution is about one order of magnitude lower than that of the Type II distribution, and Type II images also have a higher fraction above ${\cal M}=0.03$, about $1.3\%$ compared with about $0.5\%$ for Type I.
This suggests that, when the image type can be identified, Type I images may provide a cleaner reference in lensing-pair searches, or may be given a larger weight in a joint likelihood that accounts for microlensing systematics.

However, the mismatch discussed above only measures the strength of the waveform distortion.
The threshold ${\cal M}=0.03$ corresponds to a large mismatch.
Above this threshold, a signal may be missed in a matched-filter search.
However, even for signals that are not missed, with ${\cal M}<0.03$, parameter estimation can still be affected by waveform-mismatch-induced bias.
This bias is more directly related to lensing-pair searches.
Therefore, we also quantify the fraction of cluster-lensed images for which microlensing may induce a non-negligible parameter-estimation bias at LVK sensitivity.
We assess this by comparing the mismatch with the critical value for waveform indistinguishability,
\begin{equation}
{\cal M}_{\rm crit}=\chi^2_k(p) / (2 \rho^2),
\end{equation}
where $\rho$ is the GW network SNR, and $\chi^2_k(p)$ is the $p$-quantile of the chi-squared distribution with $k$ degrees of freedom.
If the mismatch is below this threshold, the microlensing-induced waveform difference is expected to be indistinguishable from statistical noise and therefore does not significantly affect parameter measurements~\citep{2010PhRvD..82b4014M, 2013PhRvD..87b4035B}.
Following~\citet{thompson2025useinterpretationsignalmodelindistinguishability}, we adopt the single-parameter case, $k=1$, as a conservative lower bound on the critical mismatch.
For a 90\% credible interval, this gives
\begin{equation}\label{eq:mismatch_crit}
{\cal M}_{\rm crit} \simeq \frac{1.35}{\rho^2},
\end{equation}
as used in~\citet{LIGOScientific:2025rsn}.
We use this threshold to identify images for which microlensing may introduce a non-negligible parameter-estimation bias.

\begin{figure}
    \centering
    \includegraphics[width=\linewidth]{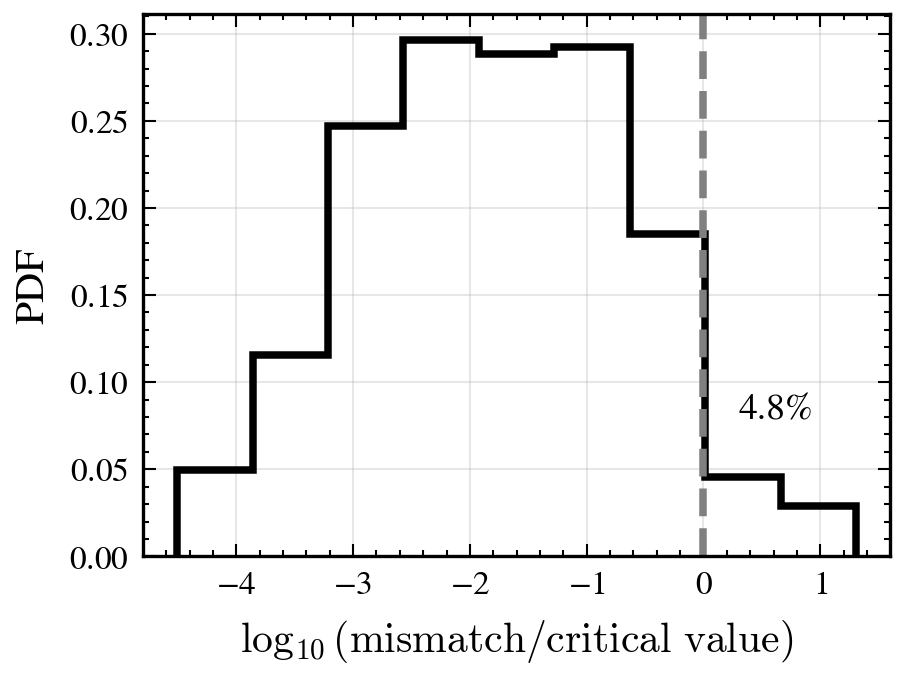}
    \caption{Distribution of $\log_{10}({\cal M}/{\cal M}_{\rm crit})$, where ${\cal M}$ is the microlensing mismatch and ${\cal M}_{\rm crit}=1.35/\rho^2$ \shan{is the mismatch threshold above which microlensing may bias parameter estimation; see Eq.~\eqref{eq:mismatch_crit}.} The vertical dashed line marks ${\cal M}/{\cal M}_{\rm crit}=1$.}
    \label{fig:mismatch_over_critical}
\end{figure}

Figure~\ref{fig:mismatch_over_critical} shows the distribution of ${\cal M}/{\cal M}_{\rm crit}$.
We find that only $4.8\%$ of cluster-lensed images have ${\cal M}/{\cal M}_{\rm crit}>1$.
Therefore, for most events lensed by clusters at LVK sensitivity, microlensing is a subdominant waveform systematic.
However, for a next-generation detector with an SNR larger by a factor of 10~\citep{Vitale_2017}, ${\cal M}_{\rm crit}$ would decrease by a factor of 100.
Applying this rescaling to the present detected sample, about $57\%$ of images would exceed the critical value, so microlensing would become an important systematic even in the cluster-lensing case.

\shan{These conclusions should be interpreted in light of the adopted microlens-population model. We assume a Chabrier IMF for living stars, a remnant population based on the initial-to-final mass relation of \citet{Spera2015}, a maximum remnant mass of $28\ M_\odot$, and a fixed remnant-to-living-star mass ratio of 0.48. Variations in the remnant mass spectrum, the total mass fraction in compact objects, or the spatial distributions of stars in the BCG, ICL, and satellite galaxies could modify the microlensing effect and the resulting mismatch distributions. In particular, a larger abundance of massive remnants could strengthen the microlensing-induced waveform distortions.}

\shan{In summary, within these astrophysical assumptions, cluster lenses produce mostly weak microlensing waveform distortions in the current LVK-sensitivity sample. Therefore, from the point of view of microlensing systematics, events lensed by clusters are promising targets for dedicated strong-lensing searches in the LVK era.}


\section{Conclusions and discussion}
\label{sec:conclusion}
\shan{As the number of detected GW events continues to grow, the first confirmed detection of a strongly lensed GW event is becoming increasingly likely~\citep{Li_2018,Xu2022,Chen:2026qtu}.
In particular, the recent lensing interpretations proposed for GW231123\_135430~\citep{o4alensing,Chan_2026,goyal2025universegw231123magnifieddiffracted,shan2025gw231123casebinarymicrolensing,2026ApJ..1003...20C,wang2026gw231123falsemassivegraviton,cheung2026diffractionlensinginterpretationgw231123astrophysical} further highlight the need to assess GW lensing effects in more realistic astrophysical environments.}
In this work, we studied the strong lensing of GWs by galaxy clusters, focusing on the time delay and magnification ratio between multiple images, together with the waveform distortion induced by microlensing. The main result is that cluster lenses do not behave as a simple extension of galaxy-scale lenses. They produce substantially longer image time delays and more complex microlensing environments due to the ICL and satellite galaxies. However, after applying the detectability cut at O4 sensitivity, their magnification-ratio distribution is similar to that of galaxy lenses.

For the time-delay distribution, we find that image pairs lensed by clusters have much longer delays than those lensed by galaxies. 
\shan{In the full simulated image-pair sample}, before applying the network-SNR cut of $\rho>8$, the cluster-lensing PDF peaks at $\Delta t\simeq110$ days, while the galaxy-lensing PDF peaks at $\Delta t\simeq19$ days.
The corresponding median delays are about $76$ days and $6.9$ days, respectively.
After applying the SNR cut at LVK O4 sensitivity, the typical delays decrease for both samples, but the detected cluster sample still has much longer delays than the detected galaxy sample: the median delay is about $15.4$ days for cluster lenses and $0.24$ days for galaxy lenses.
In addition, the detected cluster sample retains a long-delay tail, with about $9.1\%$ of image pairs having \shan{$\Delta t>365$ days} and about $3.3\%$ having $\Delta t>1000$ days.
In contrast, no detected galaxy-lensing pair reaches such long delays.
At the same time, cluster lenses can also produce very short-delay image pairs, down to about $0.01$ days, due to the contribution of substructure lensing.
Therefore, although a time-delay prior calibrated for galaxy-scale lenses is efficient for reducing the number of false candidate pairs, it can also penalize cluster-lensed pairs with naturally long delays~\citep{Diego:2021fyd}.
This motivates a targeted search for cluster-lensed GWs with a cluster-specific time-delay model, especially for cross-run event pairs.

\shan{Regarding the magnification-ratio distribution, we find that, when all simulated image pairs are considered, the cluster-lensing sample has a broader distribution than the galaxy-lensing sample because of the inclusion of member galaxies in the cluster-lensing simulation. However, after applying the detectability requirement at O4 sensitivity, the cluster- and galaxy-lensing samples have very similar magnification-ratio distributions. This suggests that differences in the magnification-ratio distributions of the two lens populations are unlikely to substantially bias the selection of strongly lensed GW candidates at O4 sensitivity. Therefore, in contrast to the time-delay prior, a magnification-ratio prior calibrated using galaxy-scale lenses appears to be less sensitive to the distinction between galaxy- and cluster-scale lens populations in the current detector era.}

\shan{Regarding the microlensing results, we find that the stellar convergence at many macro-image positions is low because these images form far from the BCG. Although the images detectable at O4 sensitivity have an average absolute macro magnification of about $8$ owing to the selection effect, this magnification is generally insufficient to compensate for the low $\kappa_\star$. Consequently, microlensing produces only small waveform distortions for most cluster-lensed images. Nevertheless, not all cluster-lensed GW images have low stellar convergences. Images located close to the BCG or satellite galaxies can have relatively high $\kappa_\star$, and the largest mismatches occur primarily for images with both high $\kappa_\star$ and high macro magnification. A higher macro magnification enlarges the micro-critical structures and, more importantly for wave-optics effects, reduces the suppression caused by the geometric time delay along the direction aligned with the shear, allowing microlenses farther from the macro-image position in the lens plane to contribute appreciably. Meanwhile, a higher stellar convergence increases the number of microlenses contributing to the waveform distortion.}

\shan{These conclusions depend on the adopted astrophysical assumptions for the microlens population and its spatial distribution. In particular, changing the maximum remnant mass from the adopted value of $28\,M_\odot$, the total mass fraction in stellar remnants, or the spatial distributions of stars in the BCG, ICL, and satellite galaxies could modify the local $\kappa_\star$ and the resulting mismatch distributions. Therefore, the quantitative microlensing fractions reported here should be interpreted within the adopted stellar-population and remnant prescriptions.}

Overall, the microlensing mismatch remains small for most detected cluster-lensed images.
We find that only about $0.8\%$ of the single-image mismatches and about $1.7\%$ of the cross-mismatches exceed the commonly used threshold ${\cal M}=0.03$.
Thus, for most cluster-lensed images, microlensing is unlikely to cause them to be missed by matched-filter searches.
We also find that Type I images have weaker microlensing effects than Type II images.
This suggests that, when the image type can be identified, Type I images may provide a cleaner reference in lensing-pair searches, or may be given a larger weight in a joint likelihood that accounts for microlensing systematics.
Finally, we quantify the level of microlensing-induced parameter-estimation bias.
At LVK sensitivity, only about $4.8\%$ of cluster-lensed images have ${\cal M}/{\cal M}_{\rm crit}>1$.
Therefore, for most events lensed by clusters at LVK sensitivity, microlensing is a subdominant waveform systematic.
However, for a next-generation detector with an SNR larger by a factor of 10, about $57\%$ of images would exceed the critical value.
Thus, microlensing will become an important systematic for future high-SNR cluster-lensing events, even though it is relatively weak in the current LVK-sensitivity sample.
This motivates the use of microlensing models in future lensing analyses, for example, through a template-free method for identifying microlensing signatures~\citep{Shan:2023ngi}, the simulation-based inference method of \citet{Su:2025xry}, or the effective \mz{description} of \citet{Zumalacarregui:2026uqs}.


Beyond their implications for GW lensing searches, cluster-lensed events may also have an advantage for identifying the host galaxy through electromagnetic observations. If a GW event is strongly lensed, its host galaxy should also be lensed by the same foreground structure. The search can therefore be limited to strong-lensing systems within the GW sky localization~\citep{Hannuksela2020,Wempe:2022zlk,Shan:2023ngi}. This may be particularly useful for cluster lenses because massive strong-lensing clusters are less common on the sky than galaxy-scale lenses. This can reduce the number of candidate lens systems that need to be checked~\citep{Hannuksela2020,Shan:2023ngi}. Once a candidate cluster is identified, the time delays and magnification ratios of the GW images can be compared with the predictions of the cluster mass model, which may help to further identify the lensed host galaxy~\citep{Smith_2018,Hannuksela2020,Wempe:2022zlk,Shan:2023ngi}.

Confirmed cluster-lensed GWs may also provide a useful way to probe additional populations of compact objects. In many of our simulated systems, the macroimages are located far from the BCG and massive cluster members. The stellar convergence is therefore low, and the microlensing effects produced by stars and stellar remnants are usually weak. This may provide a relatively clean environment for probing compact dark matter. If a fraction of dark matter is made of compact objects such as primordial black holes (PBHs), these additional compact objects may also produce observable microlensing effects on GWs~\citep{Diego:2019rzc,Urrutia:2021qak,Urrutia:2023mtk}. Therefore, if the observed frequency-dependent microlensing distortion is much stronger than expected from the stellar population model adopted in this work, it may indicate an additional compact-lens component. However, such a signal would not uniquely imply a primordial origin, because unusually massive stellar remnants or other compact structures may produce similar effects. Further studies are therefore needed to distinguish between these possibilities.

In summary, GWs lensed by clusters differ from those lensed by galaxies, with substantially longer time delays and \mz{smaller} microlensing effects, \mz{safe to ignore at current sensitivity but crucial for next-generation detectors}. By contrast, the detected cluster- and galaxy-lensing samples have similar magnification-ratio distributions at O4 sensitivity. Together with the appreciable cluster-lensing rate, these results motivate dedicated searches for GWs lensed by clusters using a cluster-specific time-delay model. Microlensing systematics need to be explicitly accounted for primarily when the event SNR is high.


\acknowledgments{
\vspace{5pt}
\textit{Acknowledgements:
We are very grateful to Luka Vujeva and Jose Maria Ezquiaga for insightful discussions. 
X.S. and S.M. acknowledge support from the National Natural Science Foundation of China (Grant No. 12133005).
H.Y. is supported by the National Natural Science Foundation of China (Grant 12573048).
X.S. acknowledges support from Shuimu Tsinghua Scholar Program (No. 2024SM199), the China Postdoctoral Science Foundation (Certificate Number: 2025M773189) and the ERC grants GWSky (101167314).
MZ acknowledges support from grant RYC2024-049805-I, funded by MICIU/AEI/10.13039/501100011033 and by the European Social Fund Plus (ESF+), and from the Ayudas de Excelencia RYC-MaX 2024 programme of the Spanish National Research Council (CSIC). This publication has been funded within the framework of the R\&D\&I Project CEX2025-001574-S, funded by MICIU/AEI/10.13039/501100011033. The research presented in this publication falls within the research line Origin and Composition of the Universe: Astroparticles and Cosmology (Astro/Cosmo). This work is supported by ERC grant GLOW (101230608). Funded by the European Union. Views and opinions expressed are however those of the author(s) only and do not necessarily reflect those of the European Union or the European Research Council Executive Agency. Neither the European Union nor the granting authority can be held responsible for them.
This material is based upon work supported by NSF's LIGO Laboratory which is a major facility fully funded by the National Science Foundation.}
   }

\appendix


\bibliographystyle{apsrev4-2}
\bibliography{example}

\end{document}